\documentclass[sigconf]{acmart}

\usepackage[font=itshape,indentfirst=true,vskip=1pt,leftmargin=0.85\leftmargin,rightmargin=0.5\leftmargin]{quoting}
\AtBeginEnvironment{quoting}{\setlength{\parindent}{-0.55\leftmargin}}

\AtBeginDocument{%
  \providecommand\BibTeX{{%
    \normalfont B\kern-0.5em{\scshape i\kern-0.25em b}\kern-0.8em\TeX}}}

\setcopyright{acmlicensed}
\copyrightyear{2021}
\acmYear{2021}
\acmDOI{10.1145/3411764.3445252}

\acmConference[CHI '21]{CHI Conference on Human Factors in Computing Systems}{May 8--13, 2021}{Yokohama, Japan}
\acmBooktitle{CHI Conference on Human Factors in Computing Systems (CHI '21), May 8--13, 2021, Yokohama, Japan}
\acmPrice{15.00}
\acmISBN{978-1-4503-8096-6/21/05}

\begin{document}

\title[No More Handshaking]{No More Handshaking: How have COVID-19 pushed\\the expansion of computer-mediated communication\\in Japanese idol culture?}

\author{Hiromu Yakura}
\orcid{0000-0002-2558-735X}
\affiliation{
    \institution{University of Tsukuba}
    \city{Tsukuba}
    \country{Japan}
}
\email{hiromu.yakura@aist.go.jp}

% \renewcommand{\shortauthors}{Anonymous Authors}

% Custom commands
\newcommand{\tabref}[1]{Table~\ref{tab:#1}}
\newcommand{\figref}[1]{Figure~\ref{fig:#1}}
\newcommand{\secref}[1]{Section~\ref{sec:#1}}
\newcommand{\eqnref}[1]{Equation~\ref{eqn:#1}}

\begin{abstract}
In Japanese idol culture, meet-and-greet events where fans were allowed to handshake with an idol member for several seconds were regarded as its essential component until the spread of COVID-19.
Now, idol groups are struggling in the transition of such events to computer-mediated communication because these events had emphasized meeting face-to-face over communicating, as we can infer from their length of time.
I anticipated that investigating this emerging transition would provide implications because their communication has a unique characteristic that is distinct from well-studied situations, such as workplace communication and intimate relationships.
Therefore, I first conducted a quantitative survey to develop a precise understanding of the transition, and based on its results, had semi-structured interviews with idol fans about their perceptions of the transition.
The survey revealed distinctive approaches, including one where fans gathered at a venue but were isolated from the idol member by an acrylic plate and talked via a video call.
Then the interviews not only provided answers to why such an approach would be reasonable but also suggested the existence of a large gap between conventional offline events and emerging online events in their perceptions.
Based on the results, I discussed how we can develop interaction techniques to support this transition and how we can apply it to other situations outside idol culture, such as computer-mediated performing arts.
\end{abstract}




\maketitle

\section{Introduction}
\label{sec:introduction}

In Japan, idol groups---which sometimes consists of more than 10 or 20 members---are dominating music record sales \cite{BB24342917}.
Their success is achieved by their marketing strategy of giving emphasis on holding meet-and-greet events to cultivate fans' loyalty to individual members \cite{Galbraith2012}.
For example, while ``handshaking events'' (\textit{akushukai}) where fans can talk for several seconds and shake hands with an idol member in exchange for buying a single CD are the most popular form \cite{Galbraith2012,Xie2014}, ``photo-session events'' (\textit{chekikai}) where fans can take an instant photo with a member are also common \cite{10.1215/10679847-3125863}.
However, the spread of COVID-19 has made it difficult to continue these events due to the risk of infection, which bereaves idol groups of the opportunity to maintain fan loyalty.
Consequently, in an analogous manner to other fields such as sales operations \cite{HARTMANN2020101,WANG2020214} and school education \cite{Kerres2020,Goldschmidt2020}, most idol groups have migrated to computer-mediated communication by introducing new forms of events.

Here, I anticipate that examining how computer technologies have been adopted in such situations will provide new insights for the HCI community regarding the use of computer-mediated communication.
This is because idol culture has a unique characteristic, which is different from other situations that previous studies have examined, such as workplace communication \cite{DBLP:conf/chi/McGregorBSAO19,DBLP:conf/chi/AbdulgalimovKNV20} and intimate relationships \cite{DBLP:conf/chi/TuYW18,DBLP:journals/pacmhci/KimLPM19}.
That is, while workplace communication mostly considers business impact rather than connectedness and intimate relationships usually emphasize connectedness, idol groups are required to balance both because they are providing the feeling of being connected with individual members as a part of their business.
This unique aspect that is not observable in daily conversation is also supported by the fact that, in conventional meet-and-greet events, fans were allowed to communicate with idol members for only a few seconds.

Given this context, I explored the following research questions to sketch future directions of computer-mediated communication from a new perspective:
\begin{description}
    \item[RQ1] How Japanese idol groups have adopted computer-mediated communication in response to COVID-19 under the dilemma between economical validity, connectedness, and, of course, technical availability?
    \item[RQ2] How idol fans have perceived the transition to computer-mediated communication, especially about its advantages and deficiencies?
\end{description}
To answer RQ1, I first surveyed the current status quantitatively and identified two main forms of online events that serve as alternatives to meet-and-greet events as well as some unique approaches reflecting their ingenuity.
I then conducted semi-structured interviews with fans of idol groups who had experience taking part in such events, either online or offline, with regards to RQ2. 
The interviews revealed the advantages and disadvantages of the transition, some of which were common to previous discussions on the use of computer-mediated communication and others that were unique to idol culture.
Based on these results, I discussed how interaction techniques can further support this transition and what findings can be applied to other situations outside of idol culture.

\section{Related work}
\label{sec:related}

In this section, I first provide the background of Japanese idol culture and its meet-and-greet events, including the current situation with COVID-19.
I then review related work that explored the use of computer-mediated communication in specific situations to contextualize this work.
I also introduce interaction techniques for supporting performing arts in distanced situations.

\subsection{Meet-and-greet events in Japanese idol culture}
\label{sec:related-idol}

As discussed in \secref{introduction}, idol groups are at the center of the Japanese music industry, as they have dominated the sales index for years \cite{BB24342917}.
However, they do not only sing songs: Holding meet-and-greet events is also a crucial component of being an ``idol'' \cite{Galbraith2012,Xie2014}.
The pioneer of such meet-and-greet events, AKB48, have promoted themselves as ``idols that you can meet'' and strategically expanded their handshaking events to exploit fans' energy by providing great intimacy and connection with individual members \cite{Galbraith2012,doi:10.1111/jpcu.12526}.
In the handshaking events held at stadiums and exhibition centers in major cities across Japan, participating fans usually wait in line for several hours to meet individual members in booths \cite{Xie2014}.
This strategic practice of offering the opportunity of face-to-face communication is followed by most idol groups, mainly female idols \cite{VanHaecke:2020:2051-7084:77}, though their events vary in scale.

However, since the Japanese government requested the cancellation of sports and cultural events in response to the spread of COVID-19 in February 2020, idol groups have had no choice but to stop such meet-and-greet events and have been seeking ways to transition to distanced situations \cite{Omiya2020,Nozaki2020}.
Here, I acknowledge that the demand for computer-mediated communication is globally observed \cite{DBLP:journals/chb/BeaunoyerDG20,HARTMANN2020101,WANG2020214,Kerres2020,Goldschmidt2020} and not specific to idol culture.
Still, it is distinctive that idol groups had previously put emphasis on the \emph{flesh-and-blood} aspect of the meet-and-greet events to provide fans with a realistic and authentic feeling \cite{Xie2014}.
This point motivates me to investigate how idol groups have adopted computer-mediated communication technologies, which would not be optimal to deliver the flesh-and-blood aspect, during this unexpected crisis.

\subsection{Use of computer-mediated communication in specific situations}
\label{sec:related-cmc}

In this context, this work shares some interest with previous studies that have investigated the use of computer-mediated communication in specific situations, such as workplace communication \cite{DBLP:conf/chi/McGregorBSAO19,DBLP:conf/chi/AbdulgalimovKNV20}.
However, as discussed in \secref{introduction}, this work could not be aligned with these studies since the aim of introducing computer-mediated communication in idol culture is to offer connectedness to fans and not to increase business impact through collaboration or knowledge-sharing.
Rather, it relates to previous studies focusing on family \cite{DBLP:conf/interact/WenKC11,DBLP:conf/hci/HanZZ19} and romantic relations \cite{DBLP:conf/chi/TuYW18,DBLP:journals/pacmhci/KimLPM19}, as they discussed how computer-mediated communication can help people feel connected to each other.
For example, it is reported that not only an instant messenger \cite{DBLP:conf/hci/HanZZ19} but also an online game \cite{DBLP:conf/interact/WenKC11} can be an important communication channel in family relations.
Still, idol groups have a different standpoint that their communication with fans is a part of their marketing strategy, and thus, they are required to balance connectedness and economical validity more severely.

This point guided me to previous studies examining the adoption of live-streaming platforms \cite{DBLP:conf/chi/TangVI16,DBLP:conf/mhci/LottridgeBWLCOR17,DBLP:conf/chi/LuXHW18,DBLP:journals/pacmhci/WangLF19}.
For instance, Tang et al. \cite{DBLP:conf/chi/TangVI16} and Lottridge et al. \cite{DBLP:conf/mhci/LottridgeBWLCOR17} depicted the personal usage of live-streaming platforms, mainly by young users in North American and European cultures.
Lu et al. \cite{DBLP:conf/chi/LuXHW18} focused on live-streamers in China, a context that is more similar to our situation, as the streamers there were also forming fan groups.
Wang et al. \cite{DBLP:journals/pacmhci/WangLF19} more closely examined Chinese live-streamers by highlighting streamer--viewer relationships supported by monetary digital gifts sent by viewers.
Actually, I have observed that Japanese idol groups often use live-streaming platforms for the current transition.
But, I also found that their usage is a bit different from what these studies reported as they are using such platforms as an alternative to individual face-to-face communication, which I describe later.

In addition, I acknowledge that some researchers in the field of fandom studies have investigated the role of computer-mediated communication, especially social networking services, in idol--fan relationships.
For example, Yan and Yang \cite{doi:10.1177/1461444820933313} analyzed the interaction between an idol and fans in Chinese social media based on the framework of parakin relationships while Sandi and Triastuti \cite{761ac28adbd940438ec2474da71faa97} discussed the case of Indonesia.
Nevertheless, Japanese idol culture has not received much research attention regarding its relation to computer technology, as it has traditionally attached importance to in-person interactions between idol members and fans.
In addition, different from the context of the above studies, the fact that idol groups have been forced to adopt computer technologies as an improvised alternative to face-to-face communication will elicit distinctive insights.
In particular, this rapid transition to computer-mediated communication is observed globally and its comparison to face-to-face communication is actively discussed in various fields, like education \cite{Muthuprasad2020} and academic conferences \cite{DBLP:conf/ict4s/ErikssonPRL20}.
Then, exploring the effect of the unique characteristic of idol culture, which I described in \secref{related-idol}, on the current situation would add a new perspective that complements these studies.

\subsection{Interaction techniques for Computer-mediated performing arts}
\label{sec:related-performance}

At the same time, I anticipate that the \emph{flesh-and-blood} aspect of conventional meet-and-greet events can be associated with performing arts.
In particular, it can be understood from the concept of \emph{aura}, which Walter Benjamin \cite{Benjamin1968} discussed in light of the introduction of a camera, as its authentic presence that is hard to reproduce by mediated technologies is the arguing point.
Since his original discussion, a long time has passed that was enough not only for the computer to be invented but also for researchers to propose various interaction technologies to address this point, which could also be applied to the current transition in idol culture.
With this motivation, I will introduce some existing techniques to support performing arts in distanced situations in a computer-mediated manner.

While live-streaming performances would be a straightforward approach, most of such techniques leverage virtual reality \cite{doi:10.1386/padm.2.1.23/1}.
This is because virtual reality is known to induce a sense of presence \cite{Slater:1997:FIV:2871096.2871098,Usoh:2000:UPQ:1246899.1246906}, which can in turn elicit authentic emotion from a user \cite{doi:10.1089/cpb.2006.9993}.
For example, Geigel et al. \cite{DBLP:journals/computer/GeigelSHJ11} presented a system that can replicate theatrical performances in a virtual space.
Kasuya et al. \cite{DBLP:conf/gamesem/KasuyaTKTMNUE19} implemented a free-viewpoint audio effect in their virtual reality-based system for distanced musical performances.
Yakura and Goto \cite{9284728} discussed methods to present realistic audience avatars in a virtual space to enhance a sense of presence during live performances.

On the other hand, some researchers argue that, in computer-mediated performance, the interactivity between performers and audience should be considered, rather than just pursue the reality of the replication, based on the idea of \emph{liveness} \cite{doi:10.1162/152028101753401767,DBLP:conf/cscw/WebbWKC16,Kim2017}.
For example, Webb et al. \cite{DBLP:conf/cscw/WebbWKC16} emphasized the sense of co-presence between performers and audience based on their interviews to expert performers and recommended constructing hybrid spaces of digital and virtual experience.
Correspondingly, Cerratto-Pargman et al. \cite{DBLP:conf/nordichi/PargmanRB14} and Wu et al. \cite{DBLP:journals/ieeemm/WuZBB17} designed interaction techniques for enabling remote audience members to actively participate in theater and music performances, respectively.

Here, I suspect that the relationship between idol members and fans can be understood analogously to that between performers and audience.
In addition, the emphasis on interactivity, which is attributed to the challenge of how to deliver aura, has some commonality with idol culture.
This point conversely suggests that investigating the current transition in idol culture has the potential to provide implications for other fields, including performing arts, which would be a discussion point later.

\section{Methods}
\label{sec:methods}

To explore the research questions presented in \secref{introduction}, I decided to use both a quantitative survey and semi-structured interviews.
The survey was prepared to develop a precise understanding of the current transition observed in idol culture, and based on its results, I conducted in-depth interviews with idol fans to unveil their perception of the transition.

\subsection{Qualitative survey}
\label{sec:methods-survey}

Correspondingly to RQ1, I first conducted a quantitative survey about how Japanese idol groups have adopted computer-mediated communication.
More specifically, I collected announcements of Japanese idol groups about holding or altering meet-and-greet events from their websites and official social media accounts.
Then, I categorized the information regarding their approaches to migrating the events into computer-mediated communication for the purpose of understanding the reasons behind their decisions.

Here, I mainly focused on female idol groups, since they were said to have more affinity with meet-and-greet events \cite{VanHaecke:2020:2051-7084:77}.
However, there are more than 3,000 female idol groups in Japan \cite{Gonohe2019}, though some of them are no longer active.
Given that the Japanese music industry makes a sharp distinction between releases from independent and major labels \cite{Stevens2012}, I narrowed the target to groups that had or planned to have a major label release.

Subsequently, I manually crawled the websites and social media accounts of each group and checked whether the group had released some announcement about meet-and-greet events between February 26\footnote{The Japanese government requested the cancellation of events on this day.} and August 25, 2020.
As a result, I collected the announcements of 53 idol groups, which had over 700 members in total.
I then listed all forms of the events, including ones held physically, from the announcements and categorized them based on their similarity.
For the events held online, I also categorized the platforms they used.

\subsection{Semi-structured Interviews}
\label{sec:methods-interviews}

The semi-structured interviews were designed to illustrate fans' perceptions of the transition of meet-and-greet events so as to answer RQ2.
Therefore, I recruited fans who had experience participating in either conventional meet-and-greet events or online alternative events of Japanese idol groups using word-of-mouth communication in local idol culture communities.
As a result, seven people (six males, one female) voluntarily participated in the interviews.
I note that the demography of the participants potentially reflects the gender homogeneity of Japanese idol culture \cite{doi:10.1111/jpcu.12526,VanHaecke:2020:2051-7084:77}.

Here, I acknowledge that the questionnaire method would be an option to reveal fans' perceptions, as Lu et al. did \cite{DBLP:conf/chi/LuXHW18}.
However, because this transition is emerging and not documented yet in the literature, I suspected that there were not enough clues to prepare a questionnaire that can lead to meaningful results.
Instead, I decided to conduct semi-structured interviews to derive in-depth perspectives from idol fans.
In addition, I note that this work prioritized reporting this living situation rather than recruiting a large number of participants, which resulted in interviews with seven participants.

The interviews were conducted via video calls in Japanese, as all participants were fluent speakers, and later translated into English.
Each interview lasted approximately 30 minutes and consisted of three main topics: ``What do you think are the advantages of this transition?''; ``What do you think are the disadvantages of this transition?''; and ``Which would you prefer, conventional offline events or recent online events, and why?''
Guided by previous studies investigating the use of computer-mediated communication \cite{DBLP:conf/interact/WenKC11,DBLP:conf/chi/TuYW18}, I analyzed the transcription of the interviews using open coding \cite{StrCor90}.
Through iterative refinement processes, I obtained several themes for each topic, which were presented later in \secref{interviews}.

\section{Results: quantitative survey}
\label{sec:survey}

\begin{figure}
    \centering
    \includegraphics[width=0.47\textwidth]{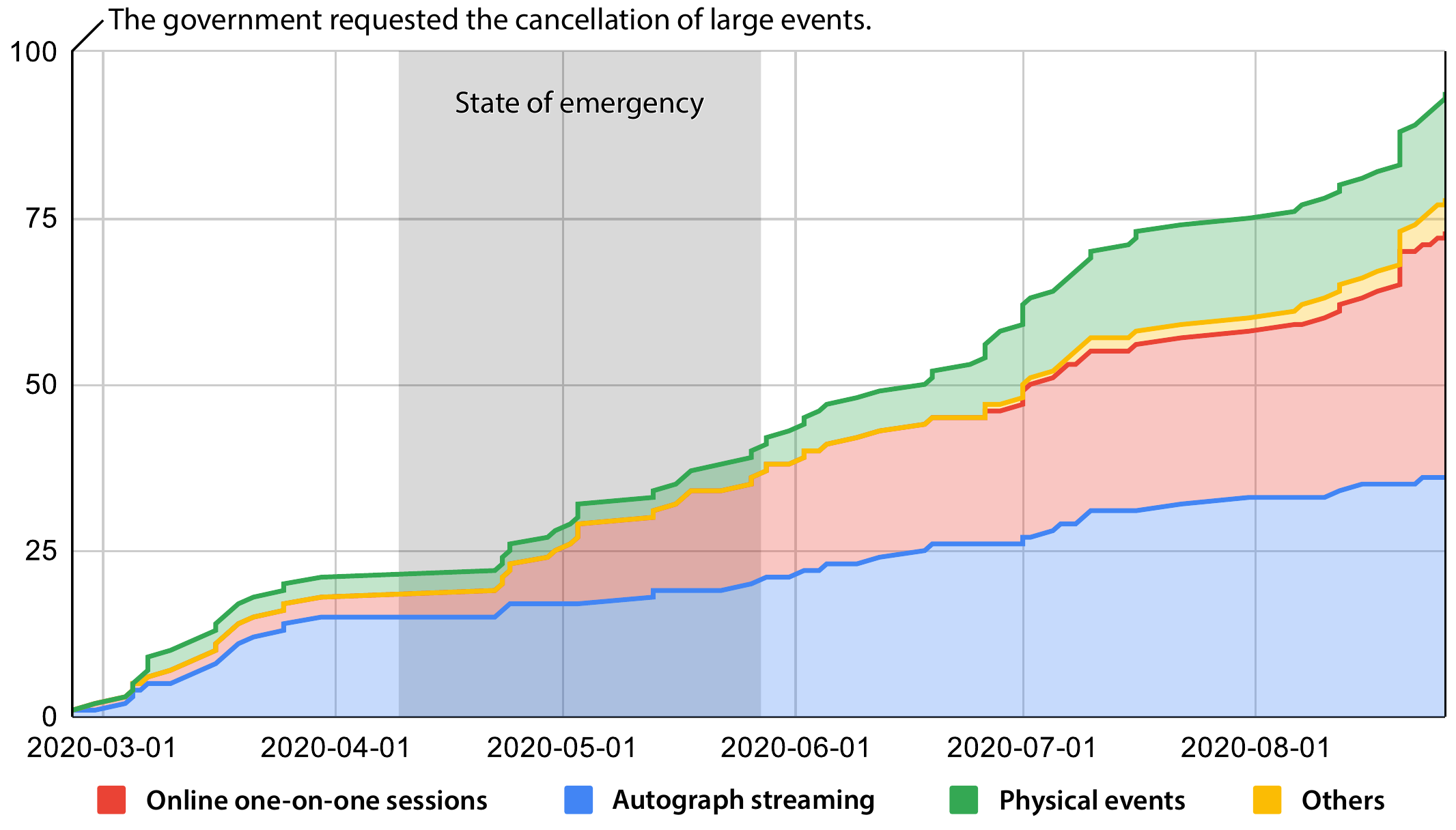}
    \caption{Timeline illustrating how Japanese idol groups have migrated their meet-and-greet events into a computer-mediated manner. Note that, since some idol groups have adopted multiple approaches, the total number in the Y-axis exceeds the number of idol groups mentioned in \secref{methods-survey}.}
    \Description[Stacked area chart illustrating the timeline of the transition]{Stacked area chart illustrating the timeline of the transition across four categories of autograph streaming, online one-on-one sessions, physical events, and others. The X-axis ranges from February 26 to August 25, 2020, and the area corresponding between April 7 and May 25 is highlighted since it was under the state of emergency.}
    \label{fig:chart}
\end{figure}

At first, I found that 17 groups (32\%) had already announced that they would resume physical meet-and-greet events as of August 25.
However, none of them involved handshaking; rather, they intended to implement an acrylic plate so that members could talk with fans without physical contact.
Some groups had planned to hold an autographing event where their members autographed any items brought by fans, but they decided instead to give out autographed cards to avoid transmitting COVID-19 by exchanging items.

Nevertheless, only three had resumed physical events without making any transition to online events.
In other words, 50 groups (94\%) had announced holding online events as an alternative to physical meet-and-greet events.
I determined that the forms of the alternative events could be divided into two major categories: autograph streaming and online one-on-one sessions.

\subsection{Autograph streaming}
\label{sec:survey-autograph}

The most popular approach, adopted by 36 groups (68\%), was streaming sessions where idol members wrote autographs in real time.
More specifically, these idol groups sell a special version of their CDs or goods in an online shop, and fans who purchase items can obtain products accompanied by an autograph with their name written by a member they specify.
Moreover, the member streams while they are writing autographs so that the fans can see the special moment when the member pronounces and writes their names, which contributes to nurturing their loyalty.
In an extreme case I found, an idol member broadcast a total of about six hours over two days to write hundreds of autographs.

This approach was initially introduced in 2014 \cite{Owada2016} to offer opportunities to interact with idol members to fans who do not live near major cities and have difficulty attending physical events.
However, it has become much more popular since the spread of COVID-19.
I postulate that the reason why this approach has been most preferred is the ease of its adoption---that is, this approach can be realized by combining existing online shops and live-streaming platforms without an additional cost.
In fact, 35 of the 36 groups were using existing live-streaming platforms to stream themselves writing autographs, such as Youtube Live (21 groups).
This point is also supported by the fact that, as shown in \figref{chart}, the announcements made in the early days were mainly categorized as autograph streaming.

On the other hand, the major weakness of this approach is its one-sidedness, which is common to the discussions on computer-mediated performing arts in \secref{related-performance}.
Especially, different from live-streamers mentioned in \secref{related-cmc}, it would be hard for idol groups to avoid the comparison with traditional handshaking events.
This point might have led them to adopt the other approach: online one-on-one sessions.

\subsection{Online one-on-one sessions}
\label{sec:survey-1on1}

This approach has also been adopted by 36 groups (68\%), 23 of whom were using both one-on-one sessions and autograph streaming.
In this approach, fans who purchased CDs or goods are assigned a time slot, typically 15 to 30 seconds long, during which they can talk to a member via a video call.
Most groups have introduced smartphone apps specifically developed for these one-on-one sessions: WithLIVE\footnote{\url{https://www.withlive.jp/}} is the most popular (12 groups) followed by talkport\footnote{\url{https://talkport.com/}} (seven groups).
On the other hand, nine groups were using Zoom\footnote{\url{http://zoom.us/}} by preparing a breakout room to which the idol member could connect and manually inviting each fan to the room during the assigned slot.

These idol groups are emphasizing this special experience by using the term ``one-on-one'' or ``individual talk,'' as well as replicating the interface of a direct video call in the introduced smartphone apps.
However, interestingly, they also emphasize that the conversations are monitored by staff in real time to avoid inappropriate fan behavior.
Moreover, in the case of the groups using Zoom, I found announcements that the staff was observing the conversations in the breakout room because there was no other way to monitor them.

\subsection{Other approaches}
\label{sec:survey-others}

I also found other, less widely used approaches.
For instance, three groups (6\%) adopted a hybrid of the above two approaches, in which an idol member writes an autograph while talking with a fan in an online one-on-one session.
I suspect that the reason why this hybrid approach is minor, although it can take advantage of both approaches, is its poor efficiency.
That is, since writing an autograph takes more time than just conversing, there are fewer fans that one member can meet in the same amount of time.
This point would be associated with the dilemma of idol groups between providing connectedness and maintaining economical validity, which I explained in \secref{related-idol}.

An idol group named SKE48 took another interesting approach: an on-site online one-on-one event.
The event was held in a convention center in the same manner as conventional handshaking events.
However, fans were not allowed to get closer than four meters to members and could not talk to them directly because of an acrylic plate placed between them.
Instead, fans could talk to the members via a video call using a pair of iPads set up in the venue while seeing the actual person through the plate.
Further, to avoid infection, fans were required to submit a medical questionnaire and measure their body temperature before entering the venue.

In another unique approach, an idol group named 22/7 held an on-site avatar-based meeting event.
For context, its members consist of voice actresses who provide the voice and motion capture for their corresponding anime characters, as well as perform as a human idol group.
Taking advantage of this feature, 22/7 held the avatar-based event in addition to online one-on-one sessions with the voice actress members.
In this event, fans could interact with a character avatar on a screen that was controlled in real time by the corresponding voice actress member using motion capture techniques.
Since the behavior of the fan is also transmitted in real time to the member in a separate room, they are able to feel as though they are talking directly to the character (or perhaps to the member).

\subsection{Analysis}
\label{sec:survey-recap}

So far, I have described how Japanese idol groups have made the transition to computer-mediated communication in response to COVID-19.
From these results, I can infer several trade-offs in making these decisions.

First, despite the risk of infection, meeting in the physical world still seems to be considered important, as many idol groups have rushed to resume their offline meet-and-greet events.
Also, the invention of the on-site online one-on-one event implies that computer-mediated communication has yet to provide an alternative to the essential component of conventional meet-and-greet events, which is common to previous studies that have reported a preference for face-to-face situations in various fields \cite{Muthuprasad2020,DBLP:conf/ict4s/ErikssonPRL20}.

Despite the importance of the close relationship between idol members and fans, the cost and effort of replicating this relationship online would also be a concern.
This not only suggests room for improvement in computer technologies but also opens up a new market for specialized applications, such as WithLIVE and talkport.

The risk of building overly intimate connections between fans and members, which has been discussed in the context of cultural studies \cite{doi:10.1111/jpcu.12526}, has been reflected in their usage of computer technologies.
That is, in real-world events, staff can easily intervene with fans who demonstrate suspicious behavior \cite{Kanai2016}.
On the other hand, while the inappropriate use of video calls---such as showing pornographic or hate images \cite{Secara2020}---is actively reported, idol groups seem to be struggling to create an intimate but controlled atmosphere in online events.

\section{Results: semi-structured interviews}
\label{sec:interviews}

The survey depicted the current transition of conventional meet-and-greet events in idol culture to computer-mediated communication as well as some dilemmas and trade-offs behind the transition.
As mentioned in \secref{methods}, I then conducted semi-structured interviews with idol fans based on the results.
The interviews that were designed to illustrate fans' perceptions of the transition provided many comments and findings, which were grouped into the advantages and disadvantages of the transition and their preferences between conventional events and online events.

\subsection{Findings: Fans' perceptions -- Advantages}
\label{sec:interviews-advantages}

\subsubsection{Fans can interact with idol members}

All participants mentioned that the fact that they had opportunities to interact with idol members during COVID-19 was irreplaceable, as follows:
\begin{quoting}
    \textnormal{P1:} I used to go to such events almost every weekend until COVID-19 spread, but then for several months, I did not have any means to communicate with members other than seeing updates on their social media. Considering that, it is really nice to be able to interact with them, regardless of its form.
\end{quoting}
\begin{quoting}
    \textnormal{P3:} For me, my \textit{oshi} [most favorite member] is literally my \textit{raison d'\^{e}tre}. So, the few months no events were held were like a hell. Now, even if it is through a screen, I appreciate the special moment that \textit{oshi} energizes me in one-on-one.
\end{quoting}
In relation to this, one participant mentioned the feeling of contributing to groups that such events created:
\begin{quoting}
    \textnormal{P5:} One of the reasons I have taken part in handshaking events is the satisfaction I feel from recognizing the fact that I am supporting the idol group by purchasing CDs. So I am glad to be able to contribute to the group through these events, especially considering that all groups are in a tough situation due to the cancellation of live concerts.
\end{quoting}
This comment is consistent with the observation on conventional meet-and-greet events \cite{Galbraith2012} that the feeling of supporting idol groups can be a factor in attracting fans.
In addition, when it comes to online events, this feeling can be aligned with the viewers' motivation to send a monetary gift to live-streamers \cite{DBLP:journals/pacmhci/WangLF19}, as mentioned in \secref{related-cmc}, though purchasing CDs would not directly benefit individual members.

\subsubsection{Ease of access}

Some participants commented about the ease of participating in online events compared to conventional offline events, as follows:
\begin{quoting}
    \textnormal{P2:} Since I live in the countryside, it always took a lot of travel time and transportation costs for me to take part in such events. So I am grateful to be able to interact with the members from my home.
\end{quoting}
\begin{quoting}
    \textnormal{P1:} So far, when multiple idol groups hold events on the same day, I have often wondered which one to go to. But, with the current online events, I can take part in different events on the same day.
\end{quoting}
While this point has often been discussed as a benefit of computer-mediated communication \cite{Muthuprasad2020,DBLP:conf/ict4s/ErikssonPRL20}, one participant mentioned a benefit specific to online one-on-one sessions as an alternative to handshaking events:
\begin{quoting}
    \textnormal{P5:} In handshaking events held at a large venue, it was common to stand in line for several hours to wait for my turn. In the online one-on-one event, the wait time is only about 30 minutes, and I can do other things during it.
\end{quoting}
However, I note that some participants described the lack of waiting time as degrading their participation experience, as described in \secref{interviews-disadvantages}.

\subsubsection{Conversation specific to the computer-mediated situation}

The participants who had taken part in online one-on-one sessions commented that they could bring up topics that were never mentioned in conventional handshaking events, as follows:
\begin{quoting}
    \textnormal{P4:} While we are not allowed to bring items to booths at offline events, it was an advantage of online sessions that I could talk to the member while showing my handmade cheering goods, such as a paper fan.
\end{quoting}
\begin{quoting}
    \textnormal{P2:} I was so happy when the member saw my room in the background and said it was cute.
\end{quoting}
However, the participant pointed out that this advantage was related to the freshness of the approach: ``If these online sessions continue for a while, such a conversation would not be special for me'' (P2).

\subsubsection{Lower psychological barriers to participation}

On the other hand, autograph streaming was also favored due to its welcoming atmosphere:
\begin{quoting}
    \textnormal{P3:} At handshaking events, my heart often beats so fast that I can't talk as much as I would like. In this regard, autograph streaming allows me to watch the member for a long time in a calm manner, and further, I can attach a short message to the member through the order form at the time of purchasing CDs.
\end{quoting}
\begin{quoting}
    \textnormal{P6:} Actually, I have never participated in a handshaking event because of my psychological barrier that is formed as I hear the various regulations of such events. When it comes to autograph streaming, I could feel free to participate, as what I need to do is just purchase a CD and watch a broadcast.
\end{quoting}
These comments suggested that this approach is not only an alternative to conventional meet-and-greet events but also a helpful means to bring in new fans.
Another participant mentioned an interesting aspect related to psychological barriers:
\begin{quoting}
    \textnormal{P1:} In offline events, I feel a bit guilty for joining the queues of multiple members of the same group one after another. On the other hand, in online events, I am free of such a feeling because the members would not know that I am talking to multiple members.
\end{quoting}
This comment highlights the unique contribution of computer technologies specific to this situation.
Here, I would like to point out that fans appreciate a simulation game-like experience that is present in their relations to idol members \cite{Xie2014,doi:10.1111/jpcu.12526}.
I suspect that computer technologies can further foster such an experience, as I discuss later in \secref{discussion}.

\subsection{Findings: Fans' perceptions -- Disadvantages}
\label{sec:interviews-disadvantages}

Despite these advantages, participants mentioned some drawbacks of the current transition to computer-mediated communication, which can be categorized into the following four topics.

\subsubsection{Technical difficulties}

The most actively discussed topic was technical issues faced by participants, mainly during online one-on-one sessions.
\begin{quoting}
    \textnormal{P4:} It was my fault for not preparing for participation in advance, but I realized just before that the OS version of my smartphone was outdated and not compatible with the designated app for one-on-one sessions. While I was updating the OS, I missed one of the time slots assigned to me.
\end{quoting}
\begin{quoting}
    \textnormal{P5:} My slow internet connection caused a huge delay during the one-on-one session and hindered smooth conversation, which left an unsatisfactory impression on me.
\end{quoting}
\begin{quoting}
    \textnormal{P2:} It was my first time using Zoom and I did not know that I needed to click a ``Join Audio'' button. So I wasted some time before I started talking.
\end{quoting}
I would like to note that, unlike the situation in education in the context of COVID-19 \cite{Muthuprasad2020}, participants did not mention issues related to device availability.
This could be attributed to the relatively high penetration rate of PCs and smartphones in Japan, especially in the younger generation, who compromise the majority of fans of idol groups \cite{Soumu2019}.
On the other hand, other issues caused by the requirement to provide opportunities for people of various levels of digital literacy \cite{DBLP:journals/chb/BeaunoyerDG20} remain consistent. 

\subsubsection{Specifications of the applications used}

Participants also mentioned that some specifications of the applications used for online one-on-one sessions were not satisfactory:
\begin{quoting}
    \textnormal{P4:} In offline handshaking events, when I bought multiple CDs, I could allocate successive slots, which in turn allowed me to talk to the member for more than a minute continuously. However, in the app the group used for online one-on-one sessions, I was disconnected each time I talked for 15 seconds and had to wait for the next assigned slot, though I had purchased multiple items. It was completely different for me to have a 1-minute session and to repeat a 15-second session four times.
\end{quoting}
\begin{quoting}
    \textnormal{P1:} As I have to wait in line for my turn in conventional offline events, I can anticipate how long will it take and prepare myself during it. However, in the online session, I was not braced to talk, as I was not notified of the order while I was in the so-called ``waiting room'' and suddenly dropped to the session.
\end{quoting}
% \begin{quoting}
%    \textnormal{P3:}  I participated in an one-on-one session using Zoom. While I was talking to the member in its breakout room, it showed that there were three users in the room: me, the member, and the staff. Though I knew in my head that it was analogous to conventional handshaking events, in which a staff stands next to a member while handshaking, but the display diminished my special feeling of talking to the member intimately.
% \end{quoting}
These issues can be expected to be resolved in the near future, especially in the applications specifically developed for this purpose.

\subsubsection{Less interactivity compared to face-to-face communication}

On the other hand, the quality of interactivity was a more serious concern, in particular when compared to face-to-face communication.
\begin{quoting}
    \textnormal{P5:} The handshake event I had planned to attend was canceled and an autograph streaming event was held instead. The moment a member said my nickname and the autographed CD that arrived later is certainly precious to me. Yet I think that its one-way interaction cannot be compared to a direct conversation in previous meet-and-greet events.
\end{quoting}
\begin{quoting}
    \textnormal{P1:} The idol group has live-streamed on a regular basis, so I could not feel the autograph streaming event was special. As a result, I could not stop doing other stuff while watching it.
\end{quoting}
The comparison to face-to-face communication was also mentioned regarding online one-on-one sessions:
\begin{quoting}
    \textnormal{P7:} I could not get myself excited in the online one-on-one session compared to conventional handshaking events. After all, I felt it was not as valuable as meeting face-to-face, even if I could talk for the same amount of time. In usual business meetings, I do not feel significant differences between online and offline situations, but when it comes to talking to idol members, I realized that even the strength of handshaking was an important part of the interaction.
\end{quoting}
\begin{quoting}
    \textnormal{P3:} In online sessions, because the only thing I could see was the face of the member captured through the camera, it was hard to know with certainty whether my feelings were being conveyed.
\end{quoting}
I acknowledge that the lack of interactivity has often been discussed as an issue in computer-mediated communication \cite{DBLP:conf/chi/TuYW18,Muthuprasad2020}.
On the other hand, these comments focusing on interaction at meet-and-greet events emphasize its specialness compared to other situations like business meetings.
I argue that this specialness can be associated with the idea of \emph{aura} in performing arts, which has motivated researchers to design specific interaction technologies to augment its sense, as I mentioned in \secref{related-performance}.
I will also discuss this point later in \secref{interviews-preference}.

\subsubsection{Lack of opportunities for fans to communicate with each other}

The participants also provided a new perspective that I had not anticipated, namely lack of communication between fans:
\begin{quoting}
    \textnormal{P7:} It was definitely one of the pleasures of handshaking events that I went to the venue with friends and chatted about things like which line to stand in or how the session was. This is not possible in the current form where I attend from home alone via the Internet.
\end{quoting}
\begin{quoting}
    \textnormal{P4:} For me, participating in handshaking events has included going for a drink with some fans who are familiar faces afterward. I know that it is hard to go out for drinks in this situation, but I miss it.
\end{quoting}
There were also some mentions of informal interactions between fans that were available at conventional offline events.
\begin{quoting}
    \textnormal{P7:} At previous handshaking events, I could hear another fan talking to the member while I was standing in line for my turn. Seeing the member affectionately talking to other fans raised my affinity for her, and it was funny to watch some fans performing a quick gag in front of the member.
\end{quoting}
Here, while I have referred to the sense of presence in \secref{related-performance} based on the analogous correspondence between performer--audience and idol--fan relationships, the importance of a sense of co-presence between audience members has also been pointed out with regards to performing arts in distanced situations \cite{DBLP:conf/collabtech/TarumiNMYT17}.
In this context, these comments can be understood from the same analogy though this aspect has been completely ignored in the current form of alternative events described in \secref{survey}.
Thus, the demand for improving computational approaches to support communication between fans is illustrated.

\subsection{Findings: Fans' preferences for face-to-face vs. computer-mediated communication}
\label{sec:interviews-preference}

Given these contexts, all participants showed a preference for conventional offline events, except for one who did not live near Tokyo, the capital of Japan, who commented:
\begin{quoting}
    \textnormal{P2:} So far, I had sometimes passed up on attending such events because of the cost to participate. In comparison, I prefer the current situation where I am able to attend all the events held online.
\end{quoting}
However, the participant continued:
\begin{quoting}
    \textnormal{P2:} On the other hand, it is unthinkable that we will never have the opportunity to meet the members physically at all. I want a chance to see them once a year, at least.
\end{quoting}
The others expressed that they want conventional meet-and-greet events held offline to return soon.
\begin{quoting}
    \textnormal{P1:} I believe the current online events are only a temporary alternative. I cannot wait to meet the members physically.
\end{quoting}
\begin{quoting}
    \textnormal{P6:} I was really happy when I saw that my favorite idol group announced the return of physical meet-and-greet events. Though I acknowledge the risk of infection, I immediately purchased the ticket, and I am really looking forward to meeting the members from now on.
\end{quoting}
These comments imply the reasons for the quick resumption of offline events, which I mentioned in \secref{survey}.
As for why fans preferred the conventional form, I obtained the following comments, which highlight perspectives that have not been actively discussed in previous literature on computer-mediated communication:
\begin{quoting}
    \textnormal{P3:} Every time I participated in meet-and-greet events, I appreciated the miracle of living under the same sky as my \textit{oshi} and could not help but thank Gods. Even though I know in my head that this fact does not change regardless of whether I am talking face-to-face or through a screen, I can't get this feeling in online communication.
\end{quoting}
\begin{quoting}
    \textnormal{P7:} In my opinion, the core experience of handshaking events is given by the special distance to a member; that is, it offers me a fantasy that I would be able to hug the member if I pull back my hand a little. Of course, I have never thought about actually trying to do so, but this point makes a huge difference in regards to whether I can meet the member physically.
\end{quoting}
\begin{quoting}
    \textnormal{P1:} The innovation of handshaking events is in the point that it presents the possibility that we can affect the members in some way through both a conversation and a handshake, whereas previous idol groups only allowed us to receive their performance in a one-way form. So, though it is often said that interactivity is the essence of offline meet-and-greet events, I would rather like to argue that the essence is the sense of what I could call affectability or intervenability, which is not well supported in the current online communication.
\end{quoting}
\begin{quoting}
    \textnormal{P4:} Even if the experience of participating in handshake events is replicated digitally in a very realistic way, for example by using expensive VR devices, I do not think it can compete with the on-site online one-on-one event held by SKE48. This is because the point is not reality, but the specialness of the experience that can only be available at the time between me and the member and cannot be copied at any other place or time. I feel that online communication inevitably detracts from this specialness, though it would not be an issue in usual online conversations.
\end{quoting}
These comments imply the importance of an aspect that would be peculiarly crucial to such meet-and-greet events and has not been well discussed.
That is, it is not just about reducing latency \cite{Erik2009} or enhancing the sense of presence by using virtual reality-based interaction technologies \cite{DBLP:journals/computer/GeigelSHJ11,DBLP:conf/gamesem/KasuyaTKTMNUE19,9284728} but about how to produce the specialness of a conversation in computer-mediated situations, as I have discussed in \secref{related-performance}.
Furthermore, one participant suggested a convenient solution for enhancing the online one-on-one session:
\begin{quoting}
    \textnormal{P7:} The fact that I am in the personal space of the member carries a lot of weight in handshake events. On that note, it is hard to see such closeness from the front camera in the online one-on-one session. If there is an additional camera of a third-person perspective that shows how closely and affectionately the member is talking to the front camera, I think that I would be more attracted.
\end{quoting}
Such an approach of augmenting distanced interactions by adding a third-person view camera has been proposed \cite{DBLP:conf/aughuman/KomiyamaMR17,DBLP:conf/socrob/TarunBCTSM19}, but the aim of these studies was efficient collaboration.
Considering that the third-person perception of one's own proxemic distance showed a similar effect as the first-person view \cite{HECHT2019113}, this approach of complementing the feeling of being in the member's personal space would be effective in fostering intimacy.

\section{Discussion}
\label{sec:discussion}

The qualitative survey and semi-structured interviews illustrated how Japanese idol groups have migrated their meet-and-greet events to computer-mediated communication and how their fans have perceived this change.
For now, considering the disadvantages listed in \secref{interviews-disadvantages}, the on-site online one-on-one event (see \secref{survey-others}) would be the optimal solution with the currently available technologies.
More specifically, fans would not face technical difficulties, as they do not need to set up any devices.
Moreover, the interactivity would be complemented by seeing the member who is talking to fans through an acrylic plate, which would also provide the special feeling of conversing under the same roof.
In addition, fans can communicate with each other, as they would gather at an on-site venue, although they would be required to follow social distancing restrictions.

However, the above results also imply further room for improvement in computer technologies since most fans preferred the conventional meet-and-greet events, as described in \secref{interviews-preference}.
In this section, I first discuss how we can provide better computational support for this situation by referring to previous literature on interaction techniques.
I also discuss which findings drawn from the results would provide implications for other situations outside of idol culture.

\subsection{How interaction techniques can help the transition to computer-mediated communication}
\label{sec:discussion-interaction}

One intuitive approach to enhancing the situations would be replicating the experience of handshaking without actually shaking hands.
In fact, there are several proposals for devices to replicate handshaking in distanced situations \cite{DBLP:conf/chi/NakanishiTW14,DBLP:conf/hri/BevanF15}.
For example, according to Nakanishi et al. \cite{DBLP:conf/chi/NakanishiTW14}, their robot hand that enables handshaking between distanced users in video-mediated communication improves users' intimacy. 
However, requiring such a device for all fans who connect from their homes would not be practical, considering the cost and difficulty of the setup.
Still, it could be deployed in the on-site online one-on-one event to complement the communication channel between idol members and fans who are distanced using an acrylic plate.

In addition, some researchers have suggested that simple tactile feedback can enhance the intimacy of distanced users in computer-mediated communication situations \cite{6512342,DBLP:conf/aughuman/BlumC16}.
Given that, it is possible to provide better experiences by enabling tactile interaction between idol members and fans in applications specifically designed for online one-on-one sessions.
Since most applications (e.g., WithLIVE and talkport) are deployed for smartphones, this can be realized without any external devices.

Furthermore, the participants in the semi-structured interviews highlighted the importance of \emph{intervenability}, a separate idea from interactivity.
As mentioned in \secref{interviews-advantages}, this can be aligned with discussions about Japanese idol culture that have pointed out the simulation game-like experience of being a fan of idol groups \cite{Xie2014,doi:10.1111/jpcu.12526}.
More specifically, offline meet-and-greet events have opened up a way for fans to affect an idol member as players of a simulation game can intervene in the story of a game by choosing options, which provides an immersive first-person perspective of being involved in the story, rather than just appreciating a pre-recorded story.
In other words, intervenability provides a sense of being involved in the relationship with a member from a first-person perspective.

In light of this, I suspect that the interaction techniques to enhance computer-mediated communication do not necessarily have to be used interactively.
For example, I might consider deploying the handshaking device only on the side of an idol member and allowing fans to manipulate it using buttons while they are talking to the member in an online one-on-one session.
This would resolve the cost of deploying such devices in fans' homes while maintaining intervenability.

Still, the comments in \secref{interviews-preference} depicted the existence of the large gap between conventional offline events and emerging online events in fans' perceptions.
In particular, it was suggested that fans are perplexed not only by the transition to computer-mediated communication but also by its consequent degradation of the flesh-and-blood and one-time aspects, namely, \emph{aura}.
This point teaches us that, as discussed in \secref{related-performance} with regards to performing arts, high-fidelity digital replication of a physical experience (e.g., by using VR devices) might not be a remedy, which was also supported by the comment in \secref{interviews-preference}.
Rather, we would be required to design specific interaction techniques that can enhance the special feeling, for example, by pursuing intervenability, as I have discussed above in this section.

\subsection{Implications for situations outside idol culture}
\label{sec:discussion-implication}

As I have discussed in relation to distanced performances, the findings of this work can provide implications for the role of computer technologies in performing arts.
In particular, as computer-mediated performing arts are becoming popular during COVID-19 \cite{10.1093/musqtl/gdaa007,doi:10.1080/01490400.2020.1774013}, the challenge of how to convey the \emph{aura} of the performances would become evident in the same manner as I observed in idol culture.
For example, we need to answer a question: What motivates users to watch the paid live-streaming of a classical music concert when they have the option to play an already purchased concert recording on the same screen?
The observation in this work suggests that on-site online events where audiences gather at a venue but are isolated by an acrylic plate to avoid infection while performances are conveyed digitally would be a reasonable choice for providing the entire experience of participating in---rather than merely watching---the performance.

In addition, the distinction of \emph{intervenability} from interactivity can also be applied to existing interactive technologies for computer-mediated performing arts, which I introduced in \secref{related-performance}.
More specifically, the theater performance reported by Cerratto-Pargman et al. \cite{DBLP:conf/nordichi/PargmanRB14} provided interactivity to an audience but not intervenability because its mechanism allowing the audience to post a text message to be displayed on the stage does not allow the audience to intervene in the plot of the performance.
On the other hand, Open Symphony proposed by Wu et al. \cite{DBLP:journals/ieeemm/WuZBB17} offers intervenability as an audience can control the mode of improvisational music performance by voting via a web application.
In this context, the discussion in \secref{discussion-interaction} implies that we have to design interaction techniques that can offer intervenability rather than interactivity so as to convey the aura in computer-mediated performing arts.

At the same time, developing such interaction techniques would also be beneficial in terms of accessibility.
As the participants mentioned in \secref{interviews-advantages}, the transition to computer-mediated communication enables people in various circumstances, including not only people living in the countryside but also people with disabilities, to participate in meet-and-greet events.
In fact, some idol groups have prepared a dedicated queue for people with disabilities in their conventional meet-and-greet events held physically \cite{Haruta2019}.
However, its access was limited to people who can at least visit the venue, which is common to performing arts in conventional theaters.
In this sense, designing interaction techniques to support the current transition that has expanded the access to a wider range of people would be demanded even if we could overcome this crisis.

Furthermore, our findings can also be connected to computer-mediated communication in personal situations.
For example, previous studies that have investigated the role of computer-mediated communication in intimate relationships \cite{DBLP:conf/chi/TuYW18,DBLP:journals/pacmhci/KimLPM19} have mainly regarded computers as a side channel to complement face-to-face communication.
However, during COVID-19, it is conceivable that couples may not be able to meet for long periods of time, especially if they are in different countries.
Then, computer-mediated communication is likely to become the main channel of their communication.
In such situations, how to introduce the flesh-and-blood aspect in computer-mediated communication would become much more important, as the participants emphasized in \secref{interviews}.
While the device-based interactions that we discussed in \secref{discussion-interaction} are possible solutions, we expect that simply adding another camera to provide a third-person perspective, as suggested in \secref{interviews-preference}, could enhance this aspect.

\section{Conclusion}
\label{sec:conclusion}

In this work, I investigated how Japanese idol groups have migrated their meet-and-greet events to computer-mediated communication due to the spread of COVID-19 and how idol fans have perceived the transition.
Correspondingly, I first surveyed the responses of 53 idol groups and illustrated the current status, including some unique approaches, such as the on-site online talk event.
I then conducted semi-structured interviews with idol fans and investigated how they perceived the transition, which suggested fans' preference for conventional meet-and-greet events due to their specialness with regards to their flesh-and-blood and one-time aspects.
Based on the results, I discussed how interaction techniques can support this transition, which not only directed us to the idea of \emph{aura} but also implied the importance of pursuing \emph{intervenability} rather than interactivity.
In addition, the discussion provided implications for several topics outside idol culture, such as computer-mediated performing arts and communication in intimate relationships, that are also under the influence of COVID-19.

On the other hand, this work has a limitation; that is, I conducted the interviews with a relatively small number of participants in order to report this living situation.
As mentioned in \secref{methods-interviews}, designing a questionnaire based on the above results and collecting responses from a large number of idol fans would provide more precise and general findings.
Still, I believe that the findings of this work originated from grassroots adaptation to computer-mediated communication are implicational for HCI researchers who are trying to design a new interaction technique.

\begin{acks}
This work was supported in part by JST ACT-X Grant Number JPMJAX200R, Japan.
\end{acks}

\bibliographystyle{ACM-Reference-Format}
\bibliography{acmart}

\end{document}